\documentclass[11pt]{article}
\usepackage{amsmath}
\usepackage{amsfonts}
\usepackage{amssymb}
\usepackage{graphicx}
\usepackage{bm}
\usepackage{color}
\usepackage{amscd}

\def\bea{\begin{eqnarray}}
\def\eea{\end{eqnarray}}

\begin{document}
\begin{center}
\LARGE {\bf Enhanced asymptotic $BMS_3$ algebra of the flat spacetime solutions of generalized minimal massive gravity}
\end{center}

\begin{center}
{M. R. Setare \footnote{E-mail: rezakord@ipm.ir}\hspace{1mm} ,
H. Adami \footnote{E-mail: hamed.adami@yahoo.com}\hspace{1.5mm} \\
{\small {\em  Department of Science, University of Kurdistan, Sanandaj, Iran.}}}\\

\end{center}

\begin{center}
{\bf{Abstract}}\\
We apply the new fall of conditions presented in the paper \cite{10} on asymptotically flat spacetime solutions of Chern-Simons-like theories of gravity. We show that the considered fall of conditions asymptotically solve equations of motion of generalized minimal massive gravity. We demonstrate that there exist two type of solutions, one of those is trivial and the others are non-trivial. By looking at non-trivial solutions, for asymptotically flat spacetimes in the generalized minimal massive gravity, in contrast to Einstein gravity, cosmological parameter can be non-zero. We obtain the conserved charges of the asymptotically flat spacetimes in generalized minimal massive gravity, and by introducing Fourier modes we show that the asymptotic symmetry algebra is a semidirect product of a $BMS_{3}$ algebra and two $U(1)$ current algebras. Also we verify that the $BMS_{3}$ algebra can be obtained by a contraction of the AdS$_3$ asymptotic symmetry algebra when the AdS$_3$ radius tends to infinity in the flat-space limit. Finally we find energy, angular momentum and entropy for a particular case and deduce that these quantities satisfy the first law of flat space cosmologies.
\end{center}

\section{Introduction}\label{1.0}
It is well known that the group of asymptotic symmetries of asymptotically flat space-times
 at future null infinity is the BMS group \cite{1',2',10'}. The BMS symmetry algebra in $n$ space-time dimension consists
 of the semi-direct sum of the conformal Killing vectors of a $(n-2)$-dimension sphere acting on the ideal of infinitesimal
 supertranslations \cite{4',5'}. In extension of AdS/CFT correspondence to the flat space holography,
 the BMS algebra has been investigated very much in recent years \cite{4',13',14',15',16',17',18',19',20',11,6',21', 160}. We know that the pure Einstein-–Hilbert gravity in three dimensions exhibits no propagating physical degrees of freedom  \cite{2'',3''}. So choosing appropriate conditions at the boundary is crucial in this theory. Depending on the chosen boundary conditions, this theory can lead to
completely different boundary theories. Recently Detournay and Riegler have introduced a new asymptotic boundary conditions for pure Einstein gravity in $2+1$ dimensions \cite{10}. In fact these boundary conditions are the flat
space counterpart of the enhanced asymptotic symmetry algebra of $AdS_3$ spacetimes which have been introduced by Troessaert previously in \cite{100}. They have shown that the resulting asymptotic symmetry algebra is generated by a $BMS_3$ algebra and two affine $U(1)$ current algebras. Then they have applied their boundary conditions to Topologically Massive Gravity (TMG) \cite{40'} and have shown that the presence of
the gravitational Chern-Simons term lead to the central extensions of the asymptotic symmetry
algebra. In the other hand TMG has a bulk-boundary unitarity conflict. Either the bulk or the boundary theory is non-unitary, so there is a clash between the positivity of the two Brown-Henneaux boundary central charges and the bulk energies. In order to overcome on this problem, Bergshoeff et.al, have introduced Minimal Massive Gravity (MMG) \cite{80}, which has the same minimal local structure as TMG. The MMG model has the same gravitational degree of freedom as the TMG. It seems that the single massive degree of freedom of MMG is
unitary in the bulk and gives rise to a unitary CFT on the boundary. Following this work Generalized Minimal Massive Gravity (GMMG) introduced \cite{8}. This model is realized by adding higher-derivative deformation term to the Lagrangian of MMG. As has been shown in \cite{8}, GMMG also avoids the aforementioned ``bulk-boundary unitarity clash''. Hamiltonian analysis show that the GMMG model has no Boulware-Deser ghosts and this model propagate only two physical modes. So this model is viable candidate for semi-classical limit of a unitary quantum $3D$ massive gravity.\\
In this paper we extend the work of \cite{10} and apply the boundary conditions introduced there to the Chern-Simons-like theories of gravity (CSLTG)
\cite{1,2}, and as a example we consider the GMMG model. It is one of the interesting extensions of \cite{10} which have been mentioned in the conclusion of \cite{10}.
\section{Quasi-local conserved charges in Chern-Simons-like theories of gravity}\label{2.0}
The Lagrangian 3-form of the Chern-Simons-like theories of gravity (CSLTG) is given by \cite{1}
\begin{equation}\label{1}
  L=\frac{1}{2} \tilde{g}_{rs} a^{r} \cdot da^{s}+\frac{1}{6} \tilde{f}_{rst} a^{r} \cdot a^{s} \times a^{t}.
\end{equation}
In the above Lagrangian $ a^{ra}=a^{ra}_{\hspace{3 mm} \mu} dx^{\mu} $ are Lorentz vector valued one-forms where, $r=1,...,N$ and $a$ indices refer to flavour and Lorentz indices, respectively. We should mention that, here, the wedge products of Lorentz-vector valued one-form fields are implicit. Also, $\tilde{g}_{rs}$ is a symmetric constant metric on the flavour space and $\tilde{f}_{rst}$ is a totally symmetric "flavour tensor" which are interpreted as the coupling constants. We use a 3D-vector algebra notation for Lorentz vectors in which contractions with $\eta _{ab}$ and $\varepsilon ^{abc}$ are denoted by dots and crosses, respectively \footnote{Here we consider the notation used in \cite{1}.}. It is worth saying that $a^{ra}$ is a collection of the dreibein $e^{a}$, the dualized spin-connection $\omega ^{a}$, the auxiliary field $ h^{a}_{\hspace{1.5 mm} \mu} = e^{a}_{\hspace{1.5 mm} \nu} h^{\nu}_{\hspace{1.5 mm} \mu} $ and so on. Also for all interesting CSLTG we have $\tilde{f}_{\omega rs} = \tilde{g}_{rs}$ \cite{2}.\\
The total variation of $a^{ra}$ due to a diffeomorphism generator $\xi$ is \cite{4}
\begin{equation}\label{2}
  \delta _{\xi} a^{ra} = \mathfrak{L}_{\xi} a^{ra} -\delta ^{r} _{\omega} d \chi _{\xi} ^{a} ,
\end{equation}
where $\chi _{\xi} ^{a}= \frac{1}{2} \varepsilon ^{a} _{\hspace{1.5 mm} bc} \lambda _{\xi}^{bc} $ and $\lambda _{\xi}^{bc}$ is generator of the Lorentz gauge transformations $SO(2, 1)$. Also, $ \delta ^{r} _{s} $  denotes the ordinary Kronecker delta and the Lorentz-Lie derivative along a vector field $\xi$ is denoted by $\mathfrak{L}_{\xi}$. We assume that $\xi$ may be a function of dynamical fields. In the paper \cite{3}, we have shown that quasi-local conserved charge perturbation associated with a field dependent vector field $\xi$ is given by \footnote{We denote variation with respect to dynamical fields by $\hat{\delta}$.}
\begin{equation}\label{3}
 \hat{\delta} Q ( \xi )  = \frac{1}{8 \pi G} \int_{\Sigma} \left( \tilde{g}_{rs} i_{\xi} a^{r} - \tilde{g} _{\omega s} \chi _{\xi} \right) \cdot \hat{\delta} a^{s},
\end{equation}
where $G$ denotes the Newtonian gravitational constant and $\Sigma$ is a spacelike codimension two surface. We can take an integration from \eqref{3} over the one-parameter path on the solution space \cite{5,6} and then we find that
\begin{equation}\label{4}
  Q ( \xi )  = \frac{1}{8 \pi G} \int_{0}^{1} ds \int_{\Sigma} \left( \tilde{g}_{rs} i_{\xi} a^{r} - \tilde{g} _{\omega s} \chi _{\xi} \right) \cdot \hat{\delta} a^{s},
\end{equation}
Also, we argued that the quasi-local conserved charge \eqref{4} is not only conserved for the Killing vectors which are admitted by spacetime everywhere but also it is conserved for the asymptotically Killing vectors.\\
In Ref. \cite{7}, we have found a general formula for the entropy of black hole solutions in CSLTG
\begin{equation}\label{5}
  \mathcal{S}= \frac{1}{4G} \int_{\text{Horizon}} \frac{d\phi}{\sqrt{g_{\phi \phi}}} \tilde{g}_{\omega s} a^{s}_{\phi \phi},
\end{equation}
where $\phi$ is angular coordinate and $g_{\phi \phi}$ denotes the $\phi \text{-} \phi$ component of spacetime metric $g_{\mu \nu}$.
\section{Generalized Minimal Massive Gravity}\label{3.0}
Generalized minimal massive gravity (GMMG) is an example of the Chern-Simons-like theories of gravity \cite{8}. In the GMMG, there are four flavours of one-form, $a^{r}= \{ e, \omega , h, f \}$, and the non-zero components of the flavour metric and the flavour tensor are
\begin{equation}\label{6}
\begin{split}
     & \tilde{g}_{e \omega}=-\sigma, \hspace{1 cm} \tilde{g}_{e h}=1, \hspace{1 cm} \tilde{g}_{\omega f}=-\frac{1}{m^{2}}, \hspace{1 cm} \tilde{g}_{\omega \omega}=\frac{1}{\mu}, \\
     & \tilde{f}_{e \omega \omega}=-\sigma, \hspace{1 cm} \tilde{f}_{e h \omega}=1, \hspace{1 cm} \tilde{f}_{f \omega \omega}=-\frac{1}{m^{2}}, \hspace{1 cm} \tilde{f}_{\omega \omega \omega}=\frac{1}{\mu},\\
     & \tilde{f} _{eff}= -\frac{1}{m^{2}}, \hspace{1 cm} \tilde{f}_{eee}=\Lambda_{0},\hspace{1 cm} \tilde{f}_{ehh}= \alpha .
\end{split}
\end{equation}
where $\sigma$, $\Lambda _{0}$, $\mu$, $m$ and $\alpha$ are a sign, cosmological parameter with dimension of mass squared, mass parameter of Lorentz Chern-Simons term, mass parameter of New Massive Gravity \cite{70} term and a dimensionless parameter, respectively. The equations of motion of  GMMG are \cite{8,9}
\begin{equation}\label{7}
   - \sigma R (\Omega) + (1 + \sigma \alpha ) D(\Omega) h - \frac{1}{2} \alpha (1 + \sigma \alpha ) h \times h + \frac{\Lambda _{0}}{2} e \times e - \frac{1}{2 m^{2}} f \times f  =0,
\end{equation}
\begin{equation}\label{8}
  - e \times f + \mu (1 + \sigma \alpha ) e \times h - \frac{\mu}{m^{2}} D(\Omega) f + \frac{\mu \alpha}{m^{2}} h \times f=0 ,
\end{equation}
\begin{equation}\label{9}
  R(\Omega) - \alpha D(\Omega) h + \frac{1}{2} \alpha ^{2} h \times h + e \times f =0,
\end{equation}
\begin{equation}\label{10}
  T(\Omega) = 0 ,
\end{equation}
where
\begin{equation}\label{11}
  \Omega = \omega - \alpha h
\end{equation}
is ordinary torsion-free dualized spin-connection. Also, $R(\Omega) = d \Omega + \frac{1}{2} \Omega \times \Omega$ is curvature 2-form, $T(\Omega)= D(\Omega)e$ is torsion 2-form, and $ D(\Omega) $ denotes exterior covariant derivative with respect to torsion-free dualized spin-connection.
\section{Asymptotically 2+1 dimensional flat spacetimes}\label{4.0}
In this section, we consider the following fall of conditions for asymptotically flat spacetimes in 3D
\begin{equation}\label{12}
  \begin{split}
     g_{uu} = & \mathcal{M}(\phi)+ \mathcal{O}(r^{-2}) \\
      g_{ur} = & -e^{\mathcal{A}(\phi)} + \mathcal{O}(r^{-2}) \\
      g_{u \phi}= & \mathcal{N}(u,\phi) + \mathcal{O}(r^{-1}) \\
      g_{rr}= & \mathcal{O}(r^{-2}) \\
      g_{r \phi}= & - e^{\mathcal{A}(\phi)} \mathcal{E}(u,\phi)+\mathcal{O}(r^{-1})  \\
      g_{\phi \phi}= & e^{2 \mathcal{A}(\phi)} r^{2} +\mathcal{E}(u,\phi) \left[ 2 \mathcal{N}(u,\phi) - \mathcal{M}(\phi) \mathcal{E}(u,\phi) \right]+\mathcal{O}(r^{-1})
  \end{split}
\end{equation}
with
\begin{equation}\label{13}
  \mathcal{N}(u,\phi)=\mathcal{L}(\phi)+\frac{u}{2} \partial_{\phi}\mathcal{M}(\phi) , \hspace{0.7 cm} \mathcal{E}(u,\phi) = \mathcal{B}(\phi)+u \partial_{\phi}\mathcal{A}(\phi)
\end{equation}
which have been introduced in the paper \cite{10}. In the above metric $\mathcal{M}(\phi)$, $\mathcal{A}(\phi)$, $\mathcal{B}(\phi)$ and $\mathcal{L}(\phi)$ are arbitrary functions. The metric, under transformation generated by vector field $\xi$, transforms as $\delta _{\xi} g_{\mu \nu} = \pounds _{\xi} g_{\mu \nu} $ \footnote{Where $\pounds _{\xi}$ denotes usual Lie derivative along $\xi$.}. The variation generated by the following Killing vector field preserves the boundary conditions
\begin{equation}\label{14}
  \begin{split}
     \xi^{u}= & \alpha(u,\phi)-\frac{1}{r} e^{-\mathcal{A}(\phi)} \mathcal{E}(u,\phi) \beta(u,\phi)+\mathcal{O}(r^{-2}), \\
      \xi^{r}= & r X(\phi) + e^{-\mathcal{A}(\phi)} \left[ \mathcal{E}(u,\phi) \partial_{\phi} X(\phi) - \partial_{\phi} \beta(u,\phi)\right], \\
       & + \frac{1}{r} e^{-2\mathcal{A}(\phi)}\beta(u,\phi) \left[ \mathcal{N}(u,\phi) - \mathcal{M}(\phi) \mathcal{E}(u,\phi) \right]+\mathcal{O}(r^{-2}), \\
      \xi^{\phi} = & Y(\phi) + \frac{1}{r} e^{-\mathcal{A}(\phi)} \beta(u,\phi)+\mathcal{O}(r^{-2}),
  \end{split}
\end{equation}
with
\begin{equation}\label{15}
  \alpha(u,\phi)=T(\phi)+ u \partial_{\phi}Y(\phi), \hspace{0.7 cm} \beta(u,\phi)= Z(\phi) + u \partial_{\phi} X(\phi),
\end{equation}
where $T(\phi)$, $X(\phi)$, $Y(\phi)$ and $Z(\phi)$ are arbitrary functions of $\phi$. Since $\xi$ depends on the dynamical fields so we need to introduce a modified
version of the Lie brackets. Let's consider a modified version of the Lie brackets \cite{11} (see also \cite{33})
\begin{equation}\label{16}
  \left[ \xi_{1},\xi_{2} \right]= \pounds _{\xi_{1}}\xi_{2}-\delta_{\xi_{1}}^{(g)} \xi_{2}+\delta_{\xi_{2}}^{(g)} \xi_{1},
\end{equation}
where $\xi_{1}=\xi(T_{1},X_{1},Y_{1},Z_{1})$ and $\xi_{2}=\xi(T_{2},X_{2},Y_{2},Z_{2})$. In the equation \eqref{16}, $\delta_{\xi_{1}}^{(g)} \xi_{2}$ denotes the change induced in $\xi_{2}$ due to the variation of metric $\delta _{\xi_{1}} g_{\mu \nu} = \pounds _{\xi_{1}} g_{\mu \nu} $. By substituting Eq.\eqref{14} into Eq.\eqref{16}, one finds
\begin{equation}\label{17}
  \left[ \xi_{1}, \xi_{2}\right]= \xi_{12},
\end{equation}
where $\xi_{12}=\xi(T_{12},X_{12},Y_{12},Z_{12})$, with
\begin{equation}\label{18}
  \begin{split}
     T_{12} =& Y_{1} \partial_{\phi} T_{2} - Y_{2} \partial_{\phi} T_{1} + T_{1} \partial_{\phi} Y_{2} - T_{2} \partial_{\phi} Y_{1},  \\
     X_{12}= & Y_{1} \partial_{\phi} X_{2} - Y_{2} \partial_{\phi} X_{1},\\
     Y_{12}= & Y_{1} \partial_{\phi} Y_{2} - Y_{2} \partial_{\phi} Y_{1},  \\
     Z_{12} =& Y_{1} \partial_{\phi} Z_{2} - Y_{2} \partial_{\phi} Z_{1} + T_{1} \partial_{\phi} X_{2} - T_{2} \partial_{\phi} X_{1}.
  \end{split}
\end{equation}
Under transformation generated by the Killing vector fields \eqref{14}, the arbitrary functions $\mathcal{M}(\phi)$, $\mathcal{A}(\phi)$, $\mathcal{B}(\phi)$ and $\mathcal{L}(\phi)$ , which have appeared in the metric, transform as
\begin{equation}\label{19}
  \begin{split}
   \delta_{\xi} \mathcal{M}(\phi)= & -2 \partial_{\phi} X(\phi) \partial_{\phi} \mathcal{A}(\phi) + 2 \partial_{\phi} Y(\phi) \mathcal{M}(\phi) + Y(\phi) \partial_{\phi} \mathcal{M}(\phi) \\
     & +2 \partial_{\phi}^{2} X(\phi),
\end{split}
\end{equation}
\begin{equation}\label{20}
  \delta_{\xi} \mathcal{A}(\phi)=Y(\phi) \partial_{\phi} \mathcal{A}(\phi)+ \partial_{\phi} Y(\phi)+X(\phi),
\end{equation}
\begin{equation}\label{21}
  \delta_{\xi} \mathcal{B}(\phi)= T(\phi) \partial_{\phi} \mathcal{A}(\phi) + Y(\phi) \partial_{\phi} \mathcal{B}(\phi) + Z(\phi) + \partial_{\phi} T(\phi),
\end{equation}
\begin{equation}\label{22}
  \begin{split}
     \delta_{\xi} \mathcal{L}(\phi)= & \partial_{\phi} T(\phi) \mathcal{M}(\phi)+Y(\phi) \partial_{\phi} \mathcal{L}(\phi)+2 \partial_{\phi} Y(\phi) \mathcal{L}(\phi)+\frac{1}{2} T(\phi) \partial_{\phi} \mathcal{M}(\phi)\\
       & - \partial_{\phi} Z(\phi) \partial_{\phi} \mathcal{A}(\phi)-\partial_{\phi} X(\phi) \partial_{\phi} \mathcal{B}(\phi)+\partial_{\phi}^{2} Z(\phi).
  \end{split}
\end{equation}
By introducing Fourier modes
\begin{equation}\label{23}
  \begin{split}
      \xi^{(T)}_{m}= & \xi(e^{im\phi},0,0,0), \\
      \xi^{(X)}_{m}= &  \xi(0,e^{im\phi},0,0),\\
      \xi^{(Y)}_{m}= &  \xi(0,0,e^{im\phi},0),\\
      \xi^{(Z)}_{m}= &  \xi(0,0,0,e^{im\phi}),
  \end{split}
\end{equation}
we will have
\begin{equation}\label{24}
  \begin{split}
       & i \left[ \xi^{(T)}_{m},\xi^{(T)}_{n} \right]=0, \hspace{0.5 cm} i \left[ \xi^{(T)}_{m},\xi^{(Z)}_{n} \right]=0, \hspace{0.5 cm} i \left[ \xi^{(X)}_{m},\xi^{(X)}_{n} \right]=0, \\
       & i \left[ \xi^{(X)}_{m},\xi^{(Z)}_{n} \right]=0, \hspace{0.5 cm} i \left[ \xi^{(Z)}_{m},\xi^{(Z)}_{n} \right]=0, \hspace{0.5 cm} i \left[ \xi^{(T)}_{m},\xi^{(X)}_{n} \right]=-n \xi^{(Z)}_{m+n}, \\
       & i \left[ \xi^{(X)}_{m},\xi^{(Y)}_{n} \right]=m \xi^{(X)}_{m+n}, \hspace{0.5 cm} i \left[ \xi^{(Y)}_{m},\xi^{(Z)}_{n} \right]=-n \xi^{(Z)}_{m+n}, \\
       & i \left[ \xi^{(T)}_{m},\xi^{(Y)}_{n} \right]=(m-n) \xi^{(T)}_{m+n}, \hspace{0.5 cm} i \left[ \xi^{(Y)}_{m},\xi^{(Y)}_{n} \right]=(m-n) \xi^{(Y)}_{m+n}.
  \end{split}
\end{equation}
Now we introduce the following dreibein
\begin{equation}\label{25}
  \begin{split}
       & e^{0}_{\hspace{1.5 mm} u}= r- \frac{1}{4r} \mathcal{M}(\phi)+\mathcal{O}(r^{-2}), \hspace{0.5 cm} e^{0}_{\hspace{1.5 mm} r}=\frac{1}{2r} e^{\mathcal{A}(\phi)}+ \mathcal{O}(r^{-2}),\\
       &  e^{0}_{\hspace{1.5 mm} \phi}=r \mathcal{E}(u,\phi)- \frac{1}{4r}\left[ 2 \mathcal{N}(u,\phi) - \mathcal{M}(\phi) \mathcal{E}(u,\phi) \right] +\mathcal{O}(r^{-2}),\\
       & e^{1}_{\hspace{1.5 mm} u}= \mathcal{O}(r^{-2}), \hspace{0.5 cm} e^{1}_{\hspace{1.5 mm} r}=\mathcal{O}(r^{-2}), \hspace{0.5 cm} e^{1}_{\hspace{1.5 mm} \phi}= r e^{\mathcal{A}(\phi)}+\mathcal{O}(r^{-2}),\\
       & e^{2}_{\hspace{1.5 mm} u}= -r- \frac{1}{4r} \mathcal{M}(\phi)+\mathcal{O}(r^{-2}), \hspace{0.5 cm} e^{2}_{\hspace{1.5 mm} r}=\frac{1}{2r} e^{\mathcal{A}(\phi)}+ \mathcal{O}(r^{-2}),\\
       &e^{2}_{\hspace{1.5 mm} \phi}=-r \mathcal{E}(u,\phi)- \frac{1}{4r}\left[ 2 \mathcal{N}(u,\phi) - \mathcal{M}(\phi) \mathcal{E}(u,\phi) \right] +\mathcal{O}(r^{-2}).
  \end{split}
\end{equation}
One can use the equation $g_{\mu \nu}= \eta _{ab} e^{a}_{\hspace{1.5 mm} \mu} e^{b}_{\hspace{1.5 mm} \nu}$, where $\eta _{ab}=\text{diag}(-1,1,1)$ is just the Minkowski metric, to obtain metric \eqref{12} from the dreibein \eqref{25}.
Since the Riemann curvature tensor $\mathcal{R}_{\alpha \beta \mu \nu}$ is related to the torsion-free curvature 2-form as
\begin{equation}\label{26}
  R^{a}(\Omega)=\frac{1}{2} e^{a}_{\hspace{1.5 mm} \lambda} \epsilon ^{\lambda \alpha \beta} \mathcal{R}_{\alpha \beta \mu \nu} dx^{\mu} \wedge dx^{\nu},
\end{equation}
therefore, for the given spacetime, we have
\begin{equation}\label{27}
  R(\Omega)=\mathcal{O}(r^{-2}).
\end{equation}
Now, in the context of GMMG, we consider following ansatz
\begin{equation}\label{28}
  f=Fe, \hspace{0.7 cm} h=He,
\end{equation}
where $F$ and $H$ are just two constant parameters. By substituting Eq.\eqref{27} and Eq.\eqref{28} into the equations of motion of GMMG \eqref{7}-\eqref{10}, we find that
\begin{equation}\label{29}
\begin{split}
     & \Lambda_{0}=\alpha \left( 1+ \alpha \sigma \right) H^{2} +\frac{F^{2}}{m^{2}}, \\
     & F= \mu \left( 1+ \alpha \sigma \right) H + \frac{\mu \alpha}{m^{2}}HF, \\
     & F+ \frac{1}{2} \alpha ^{2} H^{2}=0.
\end{split}
\end{equation}
Thus, the metric \eqref{12} solves equations of motion of GMMG asymptotically if $\Lambda_{0}$, $F$ and $H$ satisfy equations \eqref{29}. Equations \eqref{29} admit the following trivial solution
\begin{equation}\label{30}
  \Lambda_{0}=F=H=0.
\end{equation}
Now we consider the case in which $\alpha \neq 0$. In that case we have two non-trivial solutions
\begin{equation}\label{31}
  \begin{split}
     H_{\pm} = & \frac{m^{2}}{2 \mu \alpha} \pm \left[ \frac{m^{4}}{4 \mu ^{2} \alpha ^{2}}+\frac{m^{2}}{\alpha^{3}} \left( 1+ \alpha \sigma \right) \right]^{\frac{1}{2}}, \\
     F_{\pm} = & - \frac{1}{2} \alpha ^{2} H^{2}_{\pm}, \\
     \Lambda_{0\pm}= &\alpha H_{\pm}^{2} \left[ \left( 1+ \alpha \sigma \right) + \frac{\alpha^{3}}{4 m^{2}} H_{\pm}^{2}\right].
  \end{split}
\end{equation}
We mention that, if one consider the case in which $\alpha = 0$ then one again gets the trivial solution \eqref{30}. Thus, for asymptotically flat spacetimes in the GMMG model, in contrast to Einstein gravity, cosmological parameter could be non-zero.
\section{Conserved charges of asymptotically flat spacetimes in GMMG}\label{5.0}
One can use equations \eqref{6}, \eqref{28}, \eqref{29} to simplify expression \eqref{3} for that quasi-local conserved charge perturbation associated with a field dependent vector field $\xi$ in the GMMG model for asymptotically flat spacetimes
\begin{equation}\label{32}
\begin{split}
   \hat{\delta} Q ( \xi )  = \frac{1}{8 \pi G} \lim _{r \rightarrow \infty} \int_{0}^{2\pi} \biggl\{ & - \left( \sigma + \frac{\alpha H}{\mu} + \frac{F}{m^{2}} \right) \left[ i_{\xi} e \cdot \hat{\delta} \Omega _{\phi} + \left(i_{\xi} \Omega - \chi_{\xi} \right) \cdot \hat{\delta} e _{\phi} \right] \\
     & + \frac{1}{\mu} \left(i_{\xi} \Omega - \chi_{\xi} \right) \cdot \hat{\delta} \Omega _{\phi} \biggr\} d \phi.
\end{split}
\end{equation}
By demanding that the Lie-Lorentz derivative of $e^{a}$ becomes zero explicitly when $\xi$ is a Killing vector field, we find the following expression for $\chi_{\xi}$ \cite{7,12}
\begin{equation}\label{33}
  \chi _{\xi} ^{a} = i_{\xi} \omega ^{a} + \frac{1}{2} \varepsilon ^{a}_{\hspace{1.5 mm} bc} e^{\nu b} (i_{\xi} T^{c})_{\nu} + \frac{1}{2} \varepsilon ^{a}_{\hspace{1.5 mm} bc} e^{b \mu} e^{c \nu} \nabla _{\mu} \xi _{\nu} ,
\end{equation}
and one can show that this expression can be rewritten as \cite{13}
\begin{equation}\label{34}
  i_{\xi} \Omega - \chi _{\xi} = - \frac{1}{2} \varepsilon ^{a}_{\hspace{1.5 mm} bc} e^{b \mu} e^{c \nu} \nabla _{\mu} \xi _{\nu} .
\end{equation}
Also we remind that the torsion free spin-connection is given by
\begin{equation}\label{35}
  \Omega ^{a} _{ \hspace{1.5 mm}\mu} = \frac{1}{2} \varepsilon^{a b c} e _{b} ^{ \hspace{1.5 mm} \alpha} \nabla _{\mu} e_{c \alpha}.
\end{equation}
As we mentioned in section \ref{2.0}, one can take an integration from \eqref{32} over the one-parameter path on the solution space
to find the conserved charge corresponds to the Killing vector field \eqref{14} for dreibein \eqref{25}, then
\begin{equation}\label{36}
  Q(T,X,Y,Z)=M(T)+J(X)+L(Y)+P(Z),
\end{equation}
with
\begin{equation}\label{37}
  M(T)=-\frac{1}{16 \pi G} \left( \sigma + \frac{\alpha H}{\mu} + \frac{F}{m^{2}} \right) \int_{0}^{2\pi} T(\phi) \mathcal{M}(\phi) d\phi,
\end{equation}
\begin{equation}\label{38}
  J(X)=\frac{1}{8 \pi G} \int_{0}^{2\pi} X(\phi) \left[ \left( \sigma + \frac{\alpha H}{\mu} + \frac{F}{m^{2}} \right) \partial_{\phi} \mathcal{B}(\phi) - \frac{1}{2\mu}\partial_{\phi} \mathcal{A}(\phi) \right] d \phi,
\end{equation}
\begin{equation}\label{39}
\begin{split}
   L(Y)=-\frac{1}{8 \pi G} \int_{0}^{2\pi} Y(\phi) \biggl\{ & \left( \sigma + \frac{\alpha H}{\mu} + \frac{F}{m^{2}} \right) \mathcal{L}(\phi) \\
     & -\frac{1}{4\mu} \left[ 2 \mathcal{M}(\phi) + \left( \partial_{\phi} \mathcal{A}(\phi)\right)^{2} - 2 \partial_{\phi}^{2} \mathcal{A}(\phi) \right] \biggr\} d \phi,
\end{split}
\end{equation}
\begin{equation}\label{40}
  P(Z)= \frac{1}{8 \pi G} \left( \sigma + \frac{\alpha H}{\mu} + \frac{F}{m^{2}} \right) \int_{0}^{2\pi} Z(\phi) \partial_{\phi} \mathcal{A}(\phi) d \phi.
\end{equation}
The above surface charges display the universal property of 3D gravity that the space of solutions is dual to the asymptotic symmetry algebra.
The algebra of conserved charges can be written as \cite{14,15}
\begin{equation}\label{41}
  \left\{ Q(\xi _{1}) , Q(\xi _{2}) \right\} = Q \left(  \left[ \xi _{1} , \xi _{2} \right] \right) + \mathcal{C} \left( \xi _{1} , \xi _{2} \right)
\end{equation}
where $\mathcal{C} \left( \xi _{1} , \xi _{2} \right)$ is the central extension term. Also, the left hand side of the equation \eqref{41} can be defined by
\begin{equation}\label{42}
  \left\{ Q(\xi _{1}) , Q(\xi _{2}) \right\}= \hat{\delta} _{\xi _{2}} Q(\xi _{1}).
\end{equation}
Therefore one can find the central extension term by using the following formula
\begin{equation}\label{43}
  \mathcal{C} \left( \xi _{1} , \xi _{2} \right)= \hat{\delta} _{\xi _{2}} Q(\xi _{1}) - Q \left(  \left[ \xi _{1} , \xi _{2} \right] \right).
\end{equation}
By substituting Eq.\eqref{17}, Eqs.\eqref{19}-\eqref{22} and Eq.\eqref{36} into Eq.\eqref{43} we obtain the central extension term
\begin{equation}\label{44}
  \begin{split}
     \mathcal{C} \left( \xi _{1} , \xi _{2} \right)=& -\frac{1}{8 \pi G} \left( \sigma + \frac{\alpha H}{\mu} + \frac{F}{m^{2}} \right) \int_{0}^{2\pi} \biggl\{  \left(T_{1} \partial_{\phi}^{2} X_{2} -T_{2} \partial_{\phi}^{2} X_{1} \right)\\
       & \hspace{2 cm} +\left(Y_{1} \partial_{\phi}^{2} Z_{2} -Y_{2} \partial_{\phi}^{2} Z_{1} \right)- \left(X_{1} \partial_{\phi} Z_{2} -X_{2} \partial_{\phi} Z_{1} \right) \biggr\} d\phi\\
       & + \frac{1}{16 \pi G \mu} \int_{0}^{2\pi} \left\{ \left(Y_{1} \partial_{\phi}^{2} X_{2} -Y_{2} \partial_{\phi}^{2} X_{1} \right)-X_{1} \partial_{\phi} X_{2}-Y_{1} \partial_{\phi}^{3} Y_{2} \right\} d \phi.
  \end{split}
\end{equation}
By introducing Fourier modes
\begin{equation}\label{45}
  \begin{split}
      M_{m}= & Q(e^{im\phi},0,0,0)=M(e^{im\phi}), \\
      J_{m}= &  Q(0,e^{im\phi},0,0)=J(e^{im\phi}),\\
      L_{m}= &  Q(0,0,e^{im\phi},0)=L(e^{im\phi}),\\
      P_{m}= &  Q(0,0,0,e^{im\phi})=P(e^{im\phi}),
  \end{split}
\end{equation}
we find that
\begin{equation}\label{46}
\begin{split}
     & i \{ M_{m},M_{n}\}=0, \hspace{0.7 cm} i \{ M_{m},P_{n}\}=0, \hspace{0.7 cm} i \{ P_{m},P_{n}\}=0, \\
     & i \{ J_{m},J_{n}\}=k_{J}n \delta_{m+n,0}, \hspace{0.7 cm} i \{ J_{m},P_{n}\}= k_{P} n \delta_{m+n,0},\\
     & i \{ M_{m},J_{n}\}=-n P_{m+n} - i k_{P} n^{2} \delta_{m+n,0}, \\
     & i \{ J_{m},L_{n}\}=m J_{m+n}+i k_{J} m^{2} \delta_{m+n,0},\\
     & i \{ L_{m},P_{n}\}=-n P_{m+n}-i k_{P} n^{2} \delta_{m+n,0},\\
     & i \{ M_{m},L_{n}\}=(m-n) M_{m+n},\\
     & i \{ L_{m},L_{n}\}=(m-n) L_{m+n}- k_{J} n^{3} \delta_{m+n,0},
\end{split}
\end{equation}
where $k_{P}$ and $k_{J}$ are given as
\begin{equation}\label{47}
  k_{P}=-\frac{1}{4G} \left( \sigma + \frac{\alpha H}{\mu} + \frac{F}{m^{2}} \right) , \hspace{0.7 cm} k_{J}=\frac{1}{8G\mu}.
\end{equation}
Now we set $ \hat{M}_{m} \equiv M_{m} $, $ \hat{J}_{m} \equiv J_{m} $, $ \hat{L}_{m} \equiv L_{m} $ and $ \hat{P}_{m} \equiv P_{m} $, also we replace the Dirac brackets by commutators
$i\{, \} \rightarrow [, ]$, therefore we can rewritten equations \eqref{46} as following
\begin{equation}\label{48}
  \begin{split}
       & [ \tilde{L}_{m},\tilde{L}_{n}]=(m-n) \tilde{L}_{m+n}+ \frac{c_{L}}{12} m^{3} \delta_{m+n,0} \\
       & [ \tilde{M}_{m},\tilde{L}_{n}]=(m-n) \tilde{M}_{m+n}+ \frac{c_{M}}{12} m^{3} \delta_{m+n,0}, \hspace{0.7 cm} [ \tilde{M}_{m},\tilde{M}_{n}]=0,
  \end{split}
\end{equation}
\begin{equation}\label{49}
  \begin{split}
       &  [ \tilde{M}_{m},\hat{J}_{n}]=-n \hat{P}_{m+n}, \hspace{0.7 cm} [ \tilde{M}_{m},\hat{P}_{n}]=0,\\
       &  [ \tilde{L}_{m},\hat{J}_{n}]=-n \hat{J}_{m+n}, \hspace{0.7 cm} [ \tilde{L}_{m},\hat{P}_{n}]=-n \hat{P}_{m+n},\\
       &  [ \hat{P}_{m},\hat{P}_{n}]=0,\hspace{0.5 cm} [ \hat{J}_{m},\hat{J}_{n}]=k_{J} n \delta_{m+n,0},\hspace{0.5 cm}[ \hat{J}_{m},\hat{P}_{n}]=k_{p} n \delta_{m+n,0},
  \end{split}
\end{equation}
with
\begin{equation}\label{50}
  c_{L}=24k_{J}=\frac{3}{G\mu},\hspace{0.7 cm} c_{M}=12k_{P}=-\frac{3}{G} \left( \sigma + \frac{\alpha H}{\mu} + \frac{F}{m^{2}} \right),
\end{equation}
where we have performed a shift as
\begin{equation}\label{51}
  \tilde{M}_{m}=\hat{M}_{m}-im \hat{P}_{m},\hspace{0.7 cm} \tilde{L}_{m}=\hat{L}_{m}-im \hat{J}_{m}.
\end{equation}
The resulting asymptotic symmetry algebra \eqref{48} and \eqref{49} is a semidirect product of a $\mathfrak{ bms}_{3}$ algebra
,with central charges $c_{L}$ and $c_{M}$, and two $\mathfrak{u}(1)$ current algebras \cite{10}. If we set $\sigma = -1$, $\alpha=0$ and $m^{2} \rightarrow \infty$ the algebra \eqref{48} and \eqref{49} will be reduced to the one presented in \cite{10} for topologically massive gravity.\\
The algebra among the asymptotic conserved charges of asymptotically AdS$_{3}$ spacetimes in the context of GMMG is isomorphic to two copies of the
Virasoro algebra \cite{16}
\begin{equation}\label{52}
  \left[ \mathfrak{L}^{\pm} _{m},\mathfrak{L}^{\pm} _{n} \right]=(m-n)\mathfrak{L}^{\pm} _{m+n}+ \frac{c_{\pm}}{12} m^{3} \delta_{m+n,0}, \hspace{0.7 cm}\left[ \mathfrak{L}^{+} _{m},\mathfrak{L}^{-} _{n} \right]=0,
\end{equation}
where $c_{\pm}$ are central charges and they are given by \footnote{In Eq.\eqref{53}, $l$ is AdS$_{3}$ radius.}
\begin{equation}\label{53}
  c_{\pm}= -\frac{3l}{2 G} \left( \sigma + \frac{\alpha H}{\mu} + \frac{F}{m^{2}} \mp \frac{1}{\mu l} \right).
\end{equation}
The $BMS_3$ algebra \eqref{48} can be obtained by a contraction of the AdS$_{3}$ asymptotic symmetry algebra
\begin{equation}\label{54}
  \tilde{L}_{m}= \mathfrak{L}^{+}_{m}-\mathfrak{L}^{-}_{-m}, \hspace{0.7 cm} \tilde{M}_{m}=\frac{1}{l} \left( \mathfrak{L}^{+}_{m}+\mathfrak{L}^{-}_{-m} \right),
\end{equation}
when the AdS$_{3}$ radius tends to infinity in the flat-space limit \cite{17,18}. Then corresponding $BMS_3$ central charges in the algebra \eqref{48} become
\begin{equation}\label{55}
  c_{M}= \lim_{l \rightarrow \infty} \frac{1}{l} \left( c_{+}+c_{-}\right), \hspace{0.7 cm} c_{L}= \lim_{l \rightarrow \infty} \left( c_{+}-c_{-}\right),
\end{equation}
and it can be readily checked.
\section{Thermodynamics}\label{6.0}
We know that energy and angular momentum are conserved charges correspond to two asymptotic Killing vector fields $\partial_{u} $ and $ -\partial_{\phi} $, respectively. It can be seen that $\partial_{u} $ and $ -\partial_{\phi} $ are asymptotic Killing vector fields admitted by spactimes which behave asymptotically like \eqref{12} when we have $\mathcal{M}(\phi)=\mathcal{M}$, $\mathcal{A}(\phi)=\mathcal{A}$, $\mathcal{L}(\phi)=\mathcal{L}$ and $\mathcal{B}(\phi)=\mathcal{B}$, where $\mathcal{M}$, $\mathcal{A}$, $\mathcal{L}$ and $\mathcal{B}$ are constants. Hence, with this assumption, one can use Eq.\eqref{32} to find energy and angular momentum as following
\begin{equation}\label{56}
  \mathfrak{E}= Q(\partial_{u})=-\frac{1}{8G} \left( \sigma + \frac{\alpha H}{\mu} + \frac{F}{m^{2}} \right) \mathcal{M},
\end{equation}
\begin{equation}\label{57}
  \mathfrak{J}=Q(-\partial_{\phi})= \frac{1}{4G} \left[ \left( \sigma + \frac{\alpha H}{\mu} + \frac{F}{m^{2}} \right) \mathcal{L} - \frac{1}{2 \mu} \mathcal{M} \right],
\end{equation}
respectively. We know that cosmological horizon is located at where there we have
\begin{equation}\label{58}
  g_{uu}g_{\phi \phi}-\left( g_{u \phi}\right)^{2}=0,
\end{equation}
and then one can deduced that cosmological horizon is located at
\begin{equation}\label{59}
  r_{H}= \frac{e^{-\mathcal{A}}}{\sqrt{\mathcal{M}}} \left| \mathcal{L}- \mathcal{M} \mathcal{B} \right|.
\end{equation}
One can associate an angular velocity to the cosmological horizon as
\begin{equation}\label{60}
  \Omega _{H}= - \frac{g_{u \phi}}{g_{\phi \phi}}\biggr|_{r=r_{H}}=- \frac{\mathcal{M}}{\mathcal{L}}.
\end{equation}
Since the norm of Killing vector $\zeta = \partial_{u}+ \Omega _{H} \partial_{\phi}$ vanishes on the cosmological horizon, it seems sensible that one can associate a temperature to the cosmological horizon as
\begin{equation}\label{61}
  T_{H}= \frac{\kappa_{H}}{2\pi}
\end{equation}
where
\begin{equation}\label{62}
  \kappa_{H}= \left[ - \frac{1}{2} \nabla _{\mu} \zeta _{\nu}  \nabla ^{\mu} \zeta ^{\nu}\right]^{\frac{1}{2}}_{r=r_{H}},
\end{equation}
therefore we have
\begin{equation}\label{63}
  T_{H}=\frac{\mathcal{M}^{\frac{3}{2}}}{2 \pi \mathcal{L}}.
\end{equation}
As we have mentioned in section \ref{2.0}, one can obtain entropy by using Eq.\eqref{5}. Thus, we use Eq.\eqref{28} and Eq.\eqref{29} to simplify Eq.\eqref{5} for asymptotically flat spacetimes \eqref{12} in the context of GMMG
\begin{equation}\label{64}
  \mathcal{S}= \frac{1}{4G} \int_{r=r_{H}} \frac{d\phi}{\sqrt{g_{\phi \phi}}} \left[ -\left( \sigma + \frac{\alpha H}{\mu} + \frac{F}{m^{2}} \right) g_{\phi \phi} + \frac{1}{2 \mu} \Omega _{\phi \phi}\right].
\end{equation}
Since on the cosmological horizon we have
\begin{equation}\label{65}
  g_{\phi \phi} \bigr| _{r=r_{H}}= \frac{\mathcal{L}^{2}}{\mathcal{M}}, \hspace{0.7 cm} \Omega_{\phi \phi} \bigr| _{r=r_{H}}= \mathcal{L} ,
\end{equation}
then Eq.\eqref{64} becomes
\begin{equation}\label{66}
  \mathcal{S}= \frac{\pi}{2G} \left[ -\left( \sigma + \frac{\alpha H}{\mu} + \frac{F}{m^{2}} \right) \frac{\mathcal{L}}{\sqrt{\mathcal{M}}} + \frac{1}{ \mu} \sqrt{\mathcal{M}}\right].
\end{equation}
One can easily check that the quantities appear in  Eq.\eqref{56}, Eq.\eqref{57}, Eq.\eqref{60}, Eq.\eqref{63} and Eq.\eqref{66} satisfy the first law of thermodynamics of flat space cosmologies \cite{19} which is given by
\begin{equation}\label{67}
  \delta \mathfrak{E} = - T_{H} \delta \mathcal{S} + \Omega _{H} \delta \mathfrak{J}.
\end{equation}
It is easy to see that the obtained results \eqref{56}, \eqref{57} and \eqref{66} will be reduced to the corresponding results in topologically massive gravity case \cite{10} when we set $\sigma=-1$, $\alpha =0$ and $m^{2} \rightarrow \infty$.
\section{Conclusion}
In this paper we have applied the fall of conditions presented in the paper \cite{10} on asymptotically flat spacetime solutions of Chern-Simons-like theories of gravity.
In section \ref{2.0} we have reviewed the method of obtaining quasi-local conserved charges in Chern-Simons-like theories of gravity. In section \ref{3.0} we have considered generalized minimal massive gravity model as an example of Chern-Simons-like theories of gravity. The equations of motion of GMMG are given by \eqref{7}-\eqref{10}. In section \ref{4.0}, we have considered the fall of conditions \eqref{12} for the asymptotically flat spacetimes in three dimensions. The considered fall of conditions have preserved  by the variation generated by the asymptotic Killing vector field \eqref{14}. Since the asymptotic Killing vector field \eqref{14} depends on the dynamical fields, the algebra among the asymptotic Killing vectors is closed in the modified version of the Lie brackets \eqref{16}. We have considered the ansatz \eqref{28} and hence we have showed that the fall of conditions \eqref{12} asymptotically solve equations of motion of GMMG. We have obtained two types of solutions, one of those is trivial \eqref{30} and the others are non-trivial \eqref{31}. By looking at non-trivial solutions \eqref{31}, one can see that, for asymptotically flat spacetimes in the GMMG model, in contrast to Einstein gravity, cosmological parameter could be non-zero. In section \ref{5.0}, we have calculated conserved charge \eqref{36}, of asymptotically flat spacetimes, corresponds to the asymptotic Killing vector field \eqref{14}. By introducing Fourier modes \eqref{45}, we showed that asymptotic symmetry algebra, (see Eq.\eqref{48} and Eq.\eqref{49}) is a semidirect product of a $\mathfrak{ bms}_{3}$ algebra, with central charges $c_{L}$ and $c_{M}$, and two $U(1)$ current algebras. Also we verified that the $BMS_3$ algebra \eqref{48} can be obtained by a contraction of the AdS$_3$ asymptotic
symmetry algebra \eqref{52} when the AdS$_3$ radius tends to infinity in the flat-space limit. In section \ref{6.0}, we found energy, angular momentum and entropy for a particular case and we showed that they satisfy the first law of flat space cosmologies. All the obtained results in this paper will be reduced to the corresponding results in topologically massive gravity case \cite{10} when we set $\sigma=-1$, $\alpha =0$ and $m^{2} \rightarrow \infty$.
\section{Acknowledgments}
M. R. Setare thanks Max Riegler  and Blagoje Oblak for reading the manuscript, helpful comments and discussions.

\end{document}